# The physical meaning of synchronization and simultaneity in Special Relativity

Rodrigo de Abreu
Departamento de Física do IST, Universidade Técnica de Lisboa

Abstract

Based on two previous papers, the physical meaning of synchronization and simultaneity as is presented in Einstein's Special Relativity paper of 1905 is reconsidered. We follow Einstein's argumentation to introduce a criterium of synchronization and for the same arguments we arrive at a different criterium for synchronization. From that we conclude that simultaneity is absolute.

Introduction

On two previous papers [1, 2] we have analyzed the concepts of synchronization and simultaneity. We have shown in those two papers that Einstein's criterium of synchronization only in one system holds. We call that system the rest system. If we introduce a time *t'* at a *x'* through *x'=ct'* [2] for a system *S'* moving in relation to the rest system we obtain Lorentz Transformation with desynchronized clocks. But if we assign another *t'* with a similar relation where *c* is substituted by the one-way velocity of light we obtain a Synchronized Transformation where the clocks at *S'* are synchronized. We can pass from the Synchronized Transformation to the Lorentz Transformation through a Transformation of *t'*, trough a desynchronization. We have also shown how can we conceive a method to determine by experiment the absolute velocity, the velocity in relation to the rest system [3].
Now we are going to follow some of the physical arguments presented by Einstein [4] and we arrive to the same conclusions of the referred papers [1, 2]. Therefore we clarify the physical meaning of synchronization and simultaneity.

## 1. On simultaneity and synchronization.

Einstein affirms that for a "rest system" the velocity of light is *c*. If we know for that system the velocity of light we can synchronize the clocks at a given position by sending light between the several pairs of clocks that we can consider. If we know the position of two clocks we can calculate the time required for light to travel from a clock located at a position *A* to a clock located to a position *B*. The time for light to travel from clock at position *B* to the clock at position *A* is the same since we admit that the velocity of light is *c* in that frame. Therefore we can establish a "common time" for clocks *A* and *B* by putting clock *B* marking a time $t_B$ when the ray of light emitted at *A* at a time $t_A$, arrives at *B*.

$$t_B - t_A = \frac{r_{AB}}{c} \qquad (1)$$

where $r_{AB}$ is the distance between *A* e *B*.

If the ray of light is reflected at position B back to position A it arrives at A at time $t'_A$ with

$$t'_A - t_B = \frac{r_{BA}}{c} \qquad (2)$$

Since $r_{AB} = r_{BA}$

$$t'_A - t_B = t_B - t_A \qquad (3)$$

Or

$$t_B = \frac{t'_A + t_A}{2} \qquad (4)$$

This is what Einstein introduced as a criterium of synchronization, obvious in the rest system, the system where we assume that the velocity of light is $c$.

For another system moving in relation to the rest system the same criterium with the same physical meaning can be applied if we know the velocity of light in that system. Einstein continues the analysis of the meaning of simultaneity and synchronization [4] [Einstein, A. p. 127] considering, based on experience, the quantity

$$\frac{2\overline{AB}}{t'_A - t_A} = c \qquad (5)$$

as a universal constant (the velocity of light in empty space). The two-way velocity of light (5) is equal to the one-way velocity of light in the rest system. Of course (5) holds for the rest system since for the rest system the one-way velocity of light is also $c$. We also assume as Einstein the validity of (5) for every frame moving in relation to the rest system but we are not assuming that the one-way velocity of light in that system is also $c$. We cannot assume that [1, 2].

In short we accept two principles:

I. For every system the two-way velocity of light (5) is $c$.
II. For one system, the velocity of light one-way is c. We call that system the rest system.

At reference [2] we had concluded from these two principles that a coordinate $x'$ at $t=0$ has at the rest system the value $x_0$ with

$$x_0 = x'\sqrt{1 - \frac{v^2}{c^2}} \qquad (6)$$

where *v* is the velocity of system *S'* in relation to the rest system.
Therefore the *x'* axes is contracted in relation to the rest system *x* axes.

We also concluded from I and II that the time *t'* elapsed for the clock situated at the origin of *S'* (*x'=0*) is related to the time *t* elapsed for the clocks at the rest system by:

$$t = \frac{t'}{\sqrt{1-\frac{v^2}{c^2}}} \qquad (6a)$$

Acknowledging those results, let us follow Einstein analysis of simultaneity [4] [Einstein, A. p. 129]:

Consider a rod with length $r_{AB}$ as measured in the rest system with velocity *v* in relation to the rest system. *A* and *B* are the two ends of the rod "equipped with clocks that are synchronous with the clocks of the rest system, i.e. whose readings always correspond to the "time of the system at rest" at the locations the clocks happen to occupy; hence, the clocks are "synchronous in the rest system". We further imagine that each clock has an observer co-moving with it, and that these observers apply to the two clocks the criterium for the synchronous rate of two clocks formulated in section 1." [4]. But in section 1 the referred criterium was formulated for the clocks at the rest system and can not be extended for others systems. I and II implies that the one-way velocity of light in a system *S'* moving in relation to the rest system is not c, as we have demonstrated [1, 2]. Therefore, the referred criterium can not be used for the system of the rod.

Acknowledging that, let's continue with Einstein analysis [4] [Einstein, A. p.129]:

Let a ray of light start out of *A* at time $t_A$ and reflected from *B* at time $t_B$ arriving at *A* at time $t'_A$. If we consider the rest system, since the velocity of light in that system is *c* from postulate I, we have

$$c(t_B - t_A) = r_{AB} + v(t_B - t_A) \qquad (7)$$

$$c(t'_A - t_B) = r_{AB} - v(t'_A - t_B) \qquad (8)$$

From (7) and (8) we obtain

$$t_B - t_A = \frac{r_{AB}}{(c-v)} \qquad (9)$$

$$t'_A - t_B = \frac{r_{AB}}{(c+v)} \qquad (10)$$

And Einstein concludes:

"Observers co-moving with the rod would thus find that the two clocks do not run synchronously, while observers in the system at rest would declare them to be running synchronously" and therefore "we cannot ascribe absolute meaning to the concept of simultaneity". (If the clocks are marking the same time of the the rest system clocks how can they not be synchronized ? How can we be certain that the criterium can be applied to the system moving with the rod ? How can we be certain that the velocity of light is also *c* in that system).

From postulate I and II we cannot achieve that conclusion [1, 2]. From I and II the conclusion is the following:

The clocks moving with the rod have rhythms slower compared with the clocks at the rests system, eq. (6a). If we admit that the clocks at the ends of the rod are synchronized at an instant with the clocks of the rest system, since those clocks have a slower rhythm, for another instant those clocks are desynchronized with the clocks of the rest system. But this conclusion does not implies that the clocks at the ends of the rod are desynchronized between each other. Indeed if the clocks at the ends of the rod at an instant are synchronized with the clocks at the rest system they pass for every clock at the rest system marking both the same "time" although different of the "time" of the clocks of the rest system. The same rhythm, the same velocity, implies that. Therefore, if we have two events that are simultaneous for the rest system we conclude that are also for a system moving in relation to that system. The generalization for any other two systems is obvious.

## 2. Theory of Transformations of Coordinate and Time.

Einstein [4] [Einstein, A. p. 132] affirms that a ray of light emitted in time $t_A$ at a point $A$ of $S'$ moving with velocity $v$ in relation to the rest system $S$ and reflected at a point $B$ toward the point $A$ arrives there at time $t'_A$ with

$$t_B = \frac{1}{2}(t_A + t'_A) \qquad (11)$$

This condition (11) is the same condition (4) for the rest system. But for a system $S'$ moving with velocity $v$ to the rest system the condition for the synchronization of the clocks is other. We cannot use the velocity of light as *c* for the system $S'$.

We can synchronize the clocks at $S'$ if we know the velocity of light from $A$ to $B$ and from $B$ to $A$. We can obtain the physical synchronization of that clocks, with the physical criterium introduced by Einstein to obtain synchronization, but of course with the velocity of light in that system

$$c_{\rightarrow} = \frac{\overline{AB}}{(t_B - t_A)} \qquad (12)$$

$$c_{\leftarrow} = \frac{\overline{AB}}{(t'_A - t_A)} \qquad (13)$$

From (12) and (13) we have

$$\frac{1}{c_\leftarrow} + \frac{1}{c_\rightarrow} = \frac{t_B - t_A + t'_A - t_B}{\overline{AB}} = \frac{t'_A - t_A}{\overline{AB}} \qquad (14)$$

and from postulate (II), eq. (5), we have

$$\frac{1}{c_\leftarrow} + \frac{1}{c_\rightarrow} = \frac{2}{c} \qquad (15)$$

The postulate I implies that for the rest system eq. (15) is verified. But for $S'$ we can determine the velocity of light one-way through $x'$ axis satisfying also (15) [1, 2]

$$c_\rightarrow = \frac{c}{1 + \frac{v}{c}} \qquad (16)$$

$$c_\leftarrow = \frac{c}{1 - \frac{v}{c}} \qquad (17)$$

Indeed (16) and (17) satisfies postulate I, eq. (5), or the equivalent eq. (15).

Therefore we cannot impose the equality

$$c_\leftarrow = c_\rightarrow = c \qquad (18)$$

for another system. Only for one system, the rest system, we have the condition (18). The Einstein's method of synchronization with condition (18) only for one system holds. However we can synchronize the clocks in all systems. Since we consider $S$ as the rest system we can synchronize the clocks at $S'$ because we know the velocity of $S'$ in relation to $S$ and therefore we know the velocity of light one-way, (16) and (17). The result of that synchronization is obvious. All the clocks at $S'$ mark time $t'$, the same $t'$. That time is related with the time for the rest system through eq. (6a). But we are going to synchronize the clocks confirming the obvious result sending in the usual manner a ray of light between the two clocks. Suppose the clock at $x'=0$ at $t'=0$. From that point and instant a ray of light is emitted and absorbed at another $x'>0$ at an instant $t'$. The clocks are synchronized if

$$c_\rightarrow (t'-0) = x' \qquad (19)$$

or

$$t' = \frac{x'}{c} \frac{}{1+\frac{v}{c}} = \frac{x'}{c}(1+\frac{v}{c}) \qquad (20)$$

This condition must be equivalent to the other condition that all the clocks are marking the same instant $t'$. The ray of light emitted at $x'=0$ passes by a clock at $x$ when this clock marks $t=x/c$ since the velocity of light at $S$ is $c$. The ray of light passes by a corresponding clock at $S'$ when this clock marks $t'$ eq. (20). Indeed

$$x = x_0 + vt \qquad (21)$$

$$x = ct \qquad (22)$$

where $x_0$ is the the position of the second clock in $S$ at $t=0$ [2];
or from (21) and (22) we have

$$t = \frac{x_0}{c-v} \qquad (23)$$

$x'$ can be related with $x_0$, the position of the clock in $S$ at $t=0$ [2]. To determine that relation we send a ray of light from $x'=0$ to $x'$ and reflected back to $x'=0$. From postulate I the time to-and-fro $t'_o$ is

$$t'_o = \frac{2x'}{c} \qquad (24)$$

The time for these two events in $S$ is $t_o$

$$t_o = t_\rightarrow + t_\leftarrow = \frac{x_0}{c-v} + \frac{x_0}{c+v} = \frac{2x_0}{c} \frac{1}{1-\frac{v^2}{c^2}} \qquad (25)$$

From (6a), (24) and (25) we have

$$\frac{2x'}{c} = \frac{2x_0}{c} \frac{1}{1-\frac{v^2}{c^2}} \sqrt{1-\frac{v^2}{c^2}} \qquad (26)$$

or

$$x' = \frac{x_0}{\sqrt{1-\frac{v^2}{c^2}}} \qquad (27)$$

Therefore from (27) and (23) we have

$$t = \frac{x'\sqrt{1-\frac{v^2}{c^2}}}{c-v} = \frac{x'}{c}\frac{(1+\frac{v}{c})}{\sqrt{1-\frac{v^2}{c^2}}} \qquad (28)$$

and from (20):

$$t = \frac{t'}{\sqrt{1-\frac{v^2}{c^2}}} \qquad (29)$$

3. The Transformation with synchronized clocks (ST).

If the clocks at $S'$ are synchronized we have therefore the following transformation between $S$ and $S'$ from (6a), (21) and (27) [2]

$$x = x'\sqrt{1-\frac{v^2}{c^2}} + vt \qquad (30)$$

$$t = \frac{t'}{\sqrt{1-\frac{v^2}{c^2}}} \qquad (31)$$

or

$$x' = \frac{x-vt}{\sqrt{1-\frac{v^2}{c^2}}} \qquad (32)$$

$$t' = t\sqrt{1 - \frac{v^2}{c^2}} \qquad (33)$$

Of course we can made another mapping. When the ray of light emitted at $x'=0$, $t'=0$ passes by a clock at $x'$ we can put the clock marking $t'=x'/c$ and we obtain Lorentz Transformation [1]. Since the velocity of light at $S'$ is not $c$ the clocks of $S'$ with this mapping are no synchronized. One of the arguments usually presented to affirm that Einstein method of synchronization holds in any system is that we don't know what system is the rest system. Therefore it is often said that any system can be considered the rest system. But that affirmation does not mean that after admitting that a system is the rest system the others can also be considered as the rest system. If we choose, for the analysis, that a system is the rest system the others are not. Only by an experiment can we determine the rest system.

## 4. Relation between a Transformation with synchronized clocks (ST) and Lorentz Transformation (LT).

How can we pass from ST eq. (30) and (31) to LT. From (30) and (31) we can obtain LT if $X=x'$ and if we transform $t'$ to $T$ by

$$T = t' - \frac{v}{c^2} x' = t' - \frac{v}{c^2} X \quad (34)$$

or

$$t' = T + \frac{v}{c^2} X \qquad (35)$$

From (35) and (30) and (31) we obtain

$$t = \frac{T + \frac{v}{c^2} X}{\sqrt{1 - \frac{v^2}{c^2}}} \qquad (36)$$

$$x = X\sqrt{1 - \frac{v^2}{c^2}} + \frac{vT + \frac{v^2}{c^2} X}{\sqrt{1 - \frac{v^2}{c^2}}} \qquad (37)$$

or

$$x = \frac{X + vT}{\sqrt{1 - \frac{v^2}{c^2}}} \qquad (38)$$

We have with eq.(36) and (38) LT. We obtain LT introducing a dessynchronization of the clocks at $S'$ [1, 2].

## 5. A gedanken experiment to determine the absolute velocity.

Consider two systems $S'$ and $S''$ with velocities $v_1$ and $v_2$ in relation to a system $S$ the rest system [3]. Consider that $v_2 > v_1$. We assume that the origins of the three reference systems coincide when the clocks fixed at the origin of each system mark the same time $t=t'=t''=0$. We can calculate the "time" elapsed in $S$ for the origin of $S''$, $x''=0$, to arrive at coordinate $x'$. That "time" $t$ can be calculated by

$$x = v_2 t \qquad (39)$$

$$x = x'\sqrt{1 - \frac{v_1^2}{c^2}} + v_1 t \qquad (40)$$

and from (39) and (40)

$$t = \frac{x'\sqrt{1 - \frac{v_1^2}{c^2}}}{v_2 - v_1} \qquad (41)$$

The corresponding "time" $t'$ elapsed in $S'$ is from (6a) and (41)

$$t' = t\sqrt{1 - \frac{v_1^2}{c^2}} = \frac{x'(1 - \frac{v_1^2}{c^2})}{v_2 - v_1} \qquad (42)$$

This "time" is the "time" marked by the clock at $x'$ for the ST. We can pass from this "time" $t'$ to the "time" $T$ for the LT. From (34):

$$T = \frac{x'(1 - \frac{v_1^2}{c^2})}{v_2 - v_1} - \frac{v_1}{c^2} x' = \frac{x'(1 - \frac{v_1 v_2}{c^2})}{v_2 - v_1} \qquad (43)$$

This step is needed, since the observer in *S'* doesn't know yet his absolute velocity and, naturally, has "synchronized" his clocks, namely the clock at *x'* with the clock at the origin of *S'*, by the Einstein´s method.

We can also measure the "time" $t''$ elapsed at the clock located at $x''=0$. That time is (41):

$$t'' = t\sqrt{(1-\frac{v_2^2}{c^2})} = \frac{x'\sqrt{1-\frac{v_1^2}{c^2}}\sqrt{1-\frac{v_2^2}{c^2}}}{v_2 - v_1} \qquad (44)$$

From (43) and (44) we determine $v_1$ and $v_2$.

In synthesis: for two systems *S'* and *S''* in relative motion, i) begin by adjusting the clocks located at the origins of the reference systems for the same reading when the two origins pass by each other; ii) mark a distance *x´* at *S'*; iii) "synchronize" the clocks at *S'* with light by the Einstein´s method (in order that the clock at *x´* could be used later); iv) when the origin of *S"* passes by the mark *x'* at *S´*, read the clock at the origin of *S"* (time *t''*) and read the clock at *x´* in *S'* (time *T*); v) from (43) and (44) calculate the absolute velocities $v_1$ and $v_2$ of *S´*and *S´´*, respectively, in relation to the rest system *S*.

## Conclusion

We have shown in this paper that following the arguments used by Einstein at the article of 1905 about Special Relativity we are conducted to the same conclusions achieved at other two previous papers. Only in one system, the rest system, the velocity of light can be *c*, the two way velocity measured by experiment. The Lorentz Transformation is one possible transformation that we can obtain relating two systems. The Lorentz Transformation results from a particular assignment of time at a system *S'* moving in relation to the rest system *S*. Since that assignment is *x'=ct'* and the velocity of light in *S´* is not *c,* the clocks marking the time *t'* at different *x'* are not synchronized. We also obtain a Synchronized Transformation relating the same two systems. For that Transformation the clocks in each system are synchronized between each other for different *x'* since the assignment of time is made by the one-way speed of light. We also have shown how can we pass from one Transformation to the other by introducing another clock at a *x'* marking a time desynchronized to the clock correspondent to the Synchronized Transformation.

If we admit that we know the laws of physics in one system and if we wrote those Laws of physics in a covariant form, for example with differential forms, in other system in relative uniform movement we have the same laws, the same forms. We are assuming that. Therefore the covariance of physical laws does not depend of the system of coordinates that we choose. The ST Transformation is one possible Transformation that relates two systems in relative uniform movement. Therefore if we admit that we know a physical law in one system we have the same law in any other system independently of the use of ST or LT. The Lorentz Transformation reveals a symmetry that permits an introduction of mathematical entities that have the same form for the components of that entities but we must be aware of the physical meaning of that entities since the clocks of the Lorentz Transformation are desynchronized. For example the velocity of two systems is not reciprocal although the usual entitie "velocity" defined with LT coordinates is reciprocal.

A clarification of the physical meaning of synchronization and simultaneity has been presented and we also have shown how can we conceive the experimental determination of the absolute velocity.